# Nonconventional Quantized Hall Resistances Obtained with ν = 2 Equilibration in Epitaxial Graphene *p-n* Junctions

*Albert F. Rigosi[†,\*], Dinesh Patel[†,£], Martina Marzano[†,§,⊥], Mattias Kruskopf[†,‡], Heather M. Hill[†], Hanbyul Jin[†,‡], Jiuning Hu[†,‡], Angela R. Hight Walker[†], Massimo Ortolano[§], Luca Callegaro[⊥], Chi-Te Liang[£], and David B. Newell[†]*

[†]National Institute of Standards and Technology (NIST), Gaithersburg, MD 20899, USA

[£]Department of Physics, National Taiwan University, Taipei 10617, Taiwan

[§]Department of Electronics and Telecommunications, Politecnico di Torino, Torino 10129, Italy

[⊥]Istituto Nazionale di Ricerca Metrologica, Torino 10135, Italy

[‡]Joint Quantum Institute, University of Maryland, College Park, MD 20742, USA

---

[\*] Corresponding author: albert.rigosi@nist.gov; 301-975-6572

ABSTRACT

We have demonstrated the millimeter-scale fabrication of monolayer epitaxial graphene *p-n* junction devices using simple ultraviolet photolithography, thereby significantly reducing device processing time compared to that of electron beam lithography typically used for obtaining sharp junctions. This work presents measurements yielding nonconventional, fractional multiples of the typical quantized Hall resistance at $\nu = 2$ ($R_H \approx 12906$ Ω) that take the form: $\frac{a}{b}R_H$. Here, *a* and *b* have been observed to take on values such 1, 2, 3, and 5 to form various coefficients of $R_H$. Additionally, we provide a framework for exploring future device configurations using the LTspice circuit simulator as a guide to understand the abundance of available fractions one may be able to measure. These results support the potential for drastically simplifying device processing time and may be used for many other two-dimensional materials.

## 1. INTRODUCTION

Graphene has been extensively studied as a result of its great electrical and optical properties.[1-4] Epitaxial graphene (EG) on silicon carbide (SiC), which can be grown on the centimeter scale and is one of the many methods of synthesizing graphene, exhibits properties that render it suitable for large-scale or high-current applications such as the continued development of quantized Hall resistance (QHR) standards.[5-15] Though modern-day standards using millimeter-scale EG have been shown to have long-term electrical stability in ambient conditions,[16] these devices are, in most cases, only able to output a single value of quantized resistance ($\nu = 2$ plateau) to a degree of accuracy which warrants possible use in metrology. The corresponding value is: $\frac{1}{2}\frac{h}{e^2} = \frac{1}{2}R_K = R_H$, where $h$ is Planck's constant, $e$ is the elementary charge, and $R_K$ is the von Klitzing constant.

One milestone for graphene QHR standards would be the eventual accessibility of different resistance values that are well-quantized. One approach to reaching this goal includes creating quantum Hall arrays.[17-19] A major disadvantage to this approach is the requirement that many individual Hall bar devices be connected using a network of resistive interconnects, thereby increasing the total minimum device size and possibly lacking optimal contact resistances. The second approach involves building *p-n* junctions (*pnJ*s) that operate in the quantum Hall regime,[20-21] as has been previously demonstrated in EG with lateral dimensions on the order of 100 μm. EG *pnJ*s can be utilized to circumvent most of the technical difficulties resulting from the use of metallic contacts and multiple device interconnections. Research in developing materials for gating and preserving properties of large devices has seen limited success with amorphous boron nitride,[22] atomically-layered high-k dielectrics,[23-26] Parylene,[27-29] and hexagonal boron nitride,[30-31] whereas other materials have been more successful.[16, 32-33]

For millimeter-scale constructions, one major issue was fabricating correspondingly large *pnJ*s. One of the major challenges of mass producing such devices with more than one *pnJ* has been the required use of electron beam lithography, a costly and time-consuming technique, for the fabrication of junctions that are abrupt, with *n*-type and *p*-type regions separated by a width on the scale of several hundreds of nanometers or smaller. This scale is necessary to ensure that the *pnJ* is sharp enough for dissipationless equilibration of Landauer-Büttiker edge states.[20, 34] Junctions with too large a width, when dealing with bipolar interfaces, may effectively become resistive from non-quantization due to charge carrier values being in the neighborhood of the Dirac point.

In this work, we demonstrate how standard ultraviolet photolithography (UVP) and ZEP520A were used to build *pnJ*s that have junction widths smaller than 200 nm on millimeter-scale EG devices. Quantum Hall transport measurements were performed and simulated for various *p-n-p* devices to verify expected behaviors of the longitudinal resistances in a two-junction device.[35] Furthermore, we use the LTspice current simulator [see notes] to examine the various rearrangements of the electric potential in the device when injecting current at up to three independent sites. We find that nonconventional fractions of the typical quantized Hall resistance, $R_H$, can be measured, thus validating the simulations.

## 2. SAMPLE PREPARATION AND CHARACTERIZATION

*2.1 EG Growth and Device Fabrication*

The growth of high-quality epitaxial graphene can be found in Refs.[9,13,22,36] EG is formed Si atoms sublimate from the silicon face of SiC. Samples were grown on square SiC chips diced from on-axis 4*H*-SiC(0001) semi-insulating wafers (CREE) [see notes]. SiC chips were submerged in a 5:1 diluted solution of hydrofluoric acid and deionized water prior to the growth process. Chips were placed, silicon face down, on a polished graphite substrate (SPI Glas 22) [see notes] and processed with AZ5214E to utilize polymer-assisted sublimation growth techniques.[36] The face-down configuration promotes homogeneous growth,[9] and the annealing process was performed with a graphite-lined resistive-element furnace (Materials Research Furnaces Inc.) [see notes]. The heating and cooling rates were about 1.5 °C/s, with the growth performed in an ambient argon environment at 1900 °C.[9]

The grown EG was evaluated with confocal laser scanning and optical microscopy as an efficient way to identify large areas of successful growth.[37] Protective layers of Pd and Au are deposited on the EG to prevent organic contamination. While protected, the EG is etched into the desired device shape, with the final step being the removal of the protective layers from the Hall bar using a solution of 1:1 aqua regia to deionized water. To fabricate the *pnJ*s, completed Hall bars were functionalized with $Cr(CO)_3$ to reduce the electron density to a value close to the Dirac point and on the order of $10^{10}$ $cm^{-2}$. A S1813 photoresist spacer layer was then deposited on a region intended to be preserved as an *n* region. Finally, a 100 nm layer of PMMA/MMA and an approximately 350 nm layer of ZEP520A were deposited. The 100 nm layer was intended to be a mild protectant for EG since ZEP520A is very photoactive and known to reduce the mobility of EG when in direct contact with it.[32]

*2.2 Raman spectroscopy*

Raman spectroscopy was used to verify the behavior of the 2D (G') peak of the EG before and after the functionalization process and polymer photogating development. Spectra were collected with a Renishaw InVia micro-Raman spectrometer [see notes] using a 633 nm wavelength excitation laser source. The spot size was about 1 μm, the acquisition times were 30 s, the laser power was 1.7 mW power, and the optical path included a 50 × objective and 1200 $mm^{-1}$ grating. Rectangular Raman maps were collected in a backscattering configuration with step sizes of 20 μm in a 5 by 3 raster-style grid. To avoid the effects of polymer interference, spectra were collected through the backside of the SiC chip.[38]

*2.3 LTspice simulations*

The analog electronic circuit simulator LTspice was employed to predict the electrical behavior of the *pnJ* devices in several measurement configurations.[39-40] Interconnected *p*-type and *n*-type quantized regions compose the circuit and were modeled either as ideal clockwise (CW) or counterclockwise (CCW) *k*-terminal quantum Hall effect (QHE) elements. The terminal voltages $e_m$ and currents $j_m$ are related by $R_H j_m = e_m - e_{m-1}$ ($m = 1, ..., k$) for CW elements and $R_H j_m = e_m - e_{m+1}$ for CCW elements. To determine the circuit's behavior at the external terminals (to be labeled as *A* and *B*), only one polarity of magnetic flux density was simulated at a time. For a positive *B*-field, an *n*-doped (*p*-doped) graphene device was modeled by a CW (CCW) QHE element, whereas, when *B* is negative, a CWW (CW) QHE element was used.

## 3. RESULTS AND DISCUSSION

*3.1 Verifying the charge configuration*

An optical image of the EG device, fabricated into a Hall bar geometry and processed with Cr(CO)$_3$ and ZEP520A to establish two *pnJ*s, is shown in Figure 1 (a). The first and third regions separated by the UVP-obtained junctions were intended to be *p*-type regions, as indicated by the gray letters, whereas the *n* region is preserved by a thick S1813 photoresist spacer layer (red letter). Raman spectra of the device's 2D (G') peak were acquired and shown for the *n* and *p* regions immediately after transport measurements to verify the polarity of the regions. Since the thick photoresist layers prevented spectra acquired in the usual backscattering geometry, the setup was modified such that the excitation laser was shown through the backside of the SiC chip to enhance the quality of the 2D (G') peak.[38] For the data in Figure 1 (b), the *p* and *n* region spectra, which are averages of the map acquisitions, show 2D (G') peaks positioned at 2668.2 cm$^{-1}$ ± 2.3 cm$^{-1}$ and 2664.9 cm$^{-1}$ ± 4.1 cm$^{-1}$, respectively with corresponding full-widths at half-maxima of 79.1 cm$^{-1}$ ± 11.5 cm$^{-1}$ and 64.9 cm$^{-1}$ ± 9.3 cm$^{-1}$ (all uncertainties represent 1σ deviations).

Atomic force microscope (AFM) images, one of which is shown in Figure 1 (c), were used to determine the device's final thickness profile. The example profile in Figure 1 (d) was averaged over 1.1 μm and shows a height difference of about 1.4 μm between the *n* and *p* region. What became evident was that the *pnJ* width was not guaranteed to be sharp enough for dissipationless edge state equilibration. Therefore, careful treatment and analysis of the device's charge configuration was required to assess the viability of *pnJ*s created with UVP.

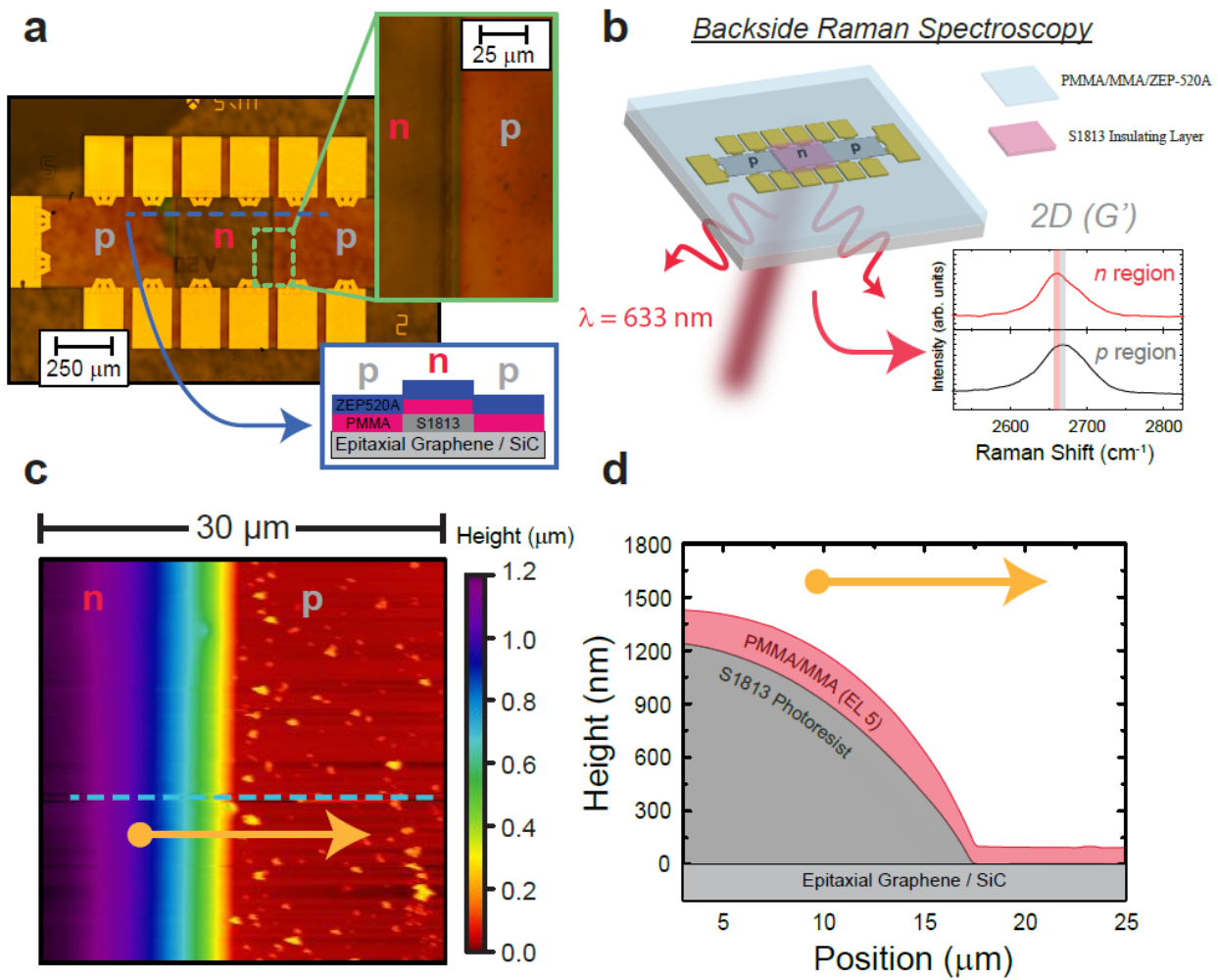

**Figure 1.** (a) The optical image of the device after processing is shown with labels indicating the intended charge polarity. A cross section of the device is also depicted for clarity. (b) An illustration of the Raman acquisition and a map-averaged 2D (G') spectrum are shown for the *n* (red) and *p* (gray) regions. The transparent red and gray bands indicate the range (for the corresponding polarity) of 2D (G') peak positions to within 1σ of the average. (c) An atomic force microscope image was acquired to gain some insight into how the boundary between the intended p and n regions formed. (d) An extracted profile prior to ZEP520A deposition is shown.

*3.2 Assessing the quality of the charge configurations and pnJs*

To assess device quality, the charge configuration of the device needed to be known and the width of the *pnJ*s needed to be estimated. It is also important to approximate how the carrier densities in the regions change with exposure to 254 nm, 17 000 µWcm$^{-2}$ UV light (distinct from the UV light used in UVP), and this is primarily done by monitoring the longitudinal resistivity in all three regions of a *p-n-p* device during a room temperature exposure, with two polarities shown in the upper panel of Figure 2 (a). For the *p* region, the expected *p*-type doping mechanism resulting from the deposition of a ZEP520A layer on the whole device persists to the point where the carrier density crosses the Dirac point. This crossing is most evident during the room temperature UV exposure when the longitudinal resistivity of the device exhibits a similar value to when the exposure was started, but instead with a negative time derivative. The S1813 successfully prevents the *n* region from becoming a *p* region, as exhibited by the flat resistivity (and electron density). For varying distances between the device and the UV lamp, as well as how the devices behave after UV exposure and without functionalization, see Supplementary Data. Though the idea of using ZEP520A as a dopant for EG has been demonstrated,[32] accessing the *p* region with that mechanism is challenging due to the intrinsic EG Fermi level pinning from the buffer layer below.[6] However, the reduction of the electron density from the order of $10^{13}$ cm$^{-2}$ to the order of $10^{10}$ cm$^{-2}$ by the presence of Cr(CO)$_3$ considerably assists the *p* region to undergo its transition.[16] It should be briefly noted that the temporary dip in resistivity near $t = 30\ 000$ s arises from another competing process to shift the carrier density, namely that of the applied heat, which as prescribed by other work, causes *n*-type doping in EG devices.[16]

Transport measurements were performed at 4 K, allowing us to determine the low-temperature longitudinal resistivity ($\rho_{xx}$), mobility ($\mu$), and electron density in EG ($n_G$), where the latter two parameters are calculated by the following respective formulas: $\mu = \frac{1}{en_G \rho_{xx}}$ and $n_G = \frac{1}{e\left(\frac{dR_{xy}}{dB}\right)}$, with *e* as the elementary charge, *B* as the magnetic flux density, and $R_{xy}$ as the Hall resistance, using SI units for all quantities. The derivative term in the denominator is calculated for magnetic flux densities under 1 T due to the linear behavior of the Hall resistance. Mobilities were measured to be on the high range of the order $10^3$ cm$^2$V$^{-1}$s$^{-1}$ and low range of the order $10^4$ cm$^2$V$^{-1}$s$^{-1}$. By mapping the change of the room temperature resistivity to that measured at 4 K, the time-dependent shift of the carrier density with UV exposure can be approximated [see Supplementary Data]. These trends are plotted in the lower panel of Figure 2 (a) for both polarities.

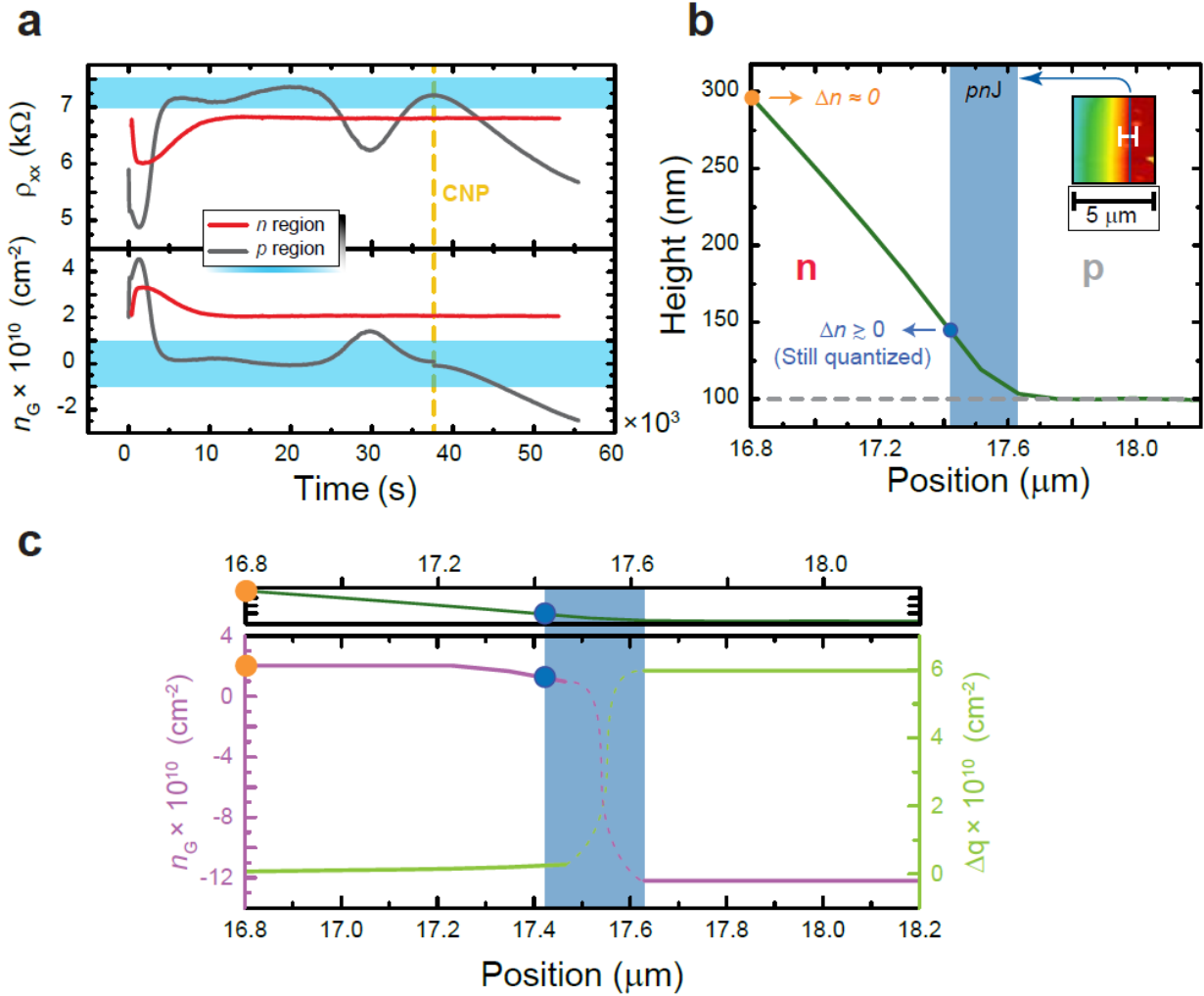

**Figure 2.** (a) The longitudinal resistivity $\rho_{xx}$ and electron density in EG $n_G$ are monitored as a function of time in the upper and lower panel, respectively, while the region is being exposed to 254 nm UV light. Surprisingly, the $Cr(CO)_3$ substantially helps the carrier density transition from *n*-type to *p*-type despite an extensive time of transient lingering close to the Dirac point. The charge neutrality point (CNP) is marked by a gold dashed line. The cyan shading approximates a range where the electrical properties of the EG would not yield quantized plateaus. (b) The AFM profile and magnified image of the *pnJ* are shown after PMMA/MMA copolymer deposition (totaling 100 nm). The green curve is taken along the white line in the inset. To validate the junction width, multiple devices with various thicknesses of S1813 were measured, as indicated

by the orange and blue dot along the profile representing the two example thicknesses of 300 nm and 42.4 nm, respectively. The shaded blue region indicates the bounds within which the carrier density is expected to switch polarity. (c) The same profile and shaded region is projected onto the calculated charge transfer $\Delta q$ to the ZEP520A layer and profile of $n_G$ (both as a function of lateral distance).

So far, our assessment of the viability of *pnJ* creation has only focused on how regions' polarities can be converted from *n*-type to *p*-type, and how these devices respond to the conversion process. More crucially, to determine the width of the *pnJ*, a capacitance model was implemented using parameters from a magnified AFM profile in Figure 2 (b) to gain insight on the amount of expected charge transfer from EG to ZEP520A in the *p* and *n* regions. For the model, we must consider the quantum capacitance of the charge transfer layers which includes EG, the buffer layer beneath, and the residual chemical doping between graphene and polymer layers above it. The general properties of the interfacial buffer layer are well-known, with its presence leading to *n*-type doping in EG.[20, 41-43] Recall that $E_F$ is the Fermi level of the EG layer, respectively, with $E_F = \hbar v_F \sqrt{\pi |n_G|} sign(n_G)$. The relationship between those two parameters and the amount of charge transferred from graphene to ZEP520A is given by:[20, 41, 44, 45]

$$\frac{C_{poly}}{e}\left(\frac{e\Delta q}{C_{poly}} - V_D\right) = \frac{C_{\gamma 2}}{C_{s2}}\left[n_G + (C_{s1} + C_{s2})\frac{E_F}{e^2}\right] \quad (1)$$

In equation (1), $\frac{e\Delta q}{C_{poly}}$ replaces the term that typically represents an electrostatic gate ($V_G$),[41] where $\Delta q$ is the amount of charge transferred. The polymer gate geometric capacitance (all capacitances are per unit area) is $C_{poly} = \frac{\epsilon_{poly}}{d_{poly}}$, where $\epsilon_{poly}$ and $d_{poly}$ are the dielectric constant and thickness of the used polymer (either S1813 or PMMA/MMA or a combination in series), respectively. $V_D$ is the voltage corresponding to the Dirac point, which is approximately 16.5 meV ($n_G = 2 \times 10^{10}$ cm$^{-2}$), as per the expected behavior of functionalized EG.[16] The summed capacitance, taking on the subscript $i = 1$ for the buffer layer and $i = 2$ for residual chemical doping, is $C_{si} = \left(\frac{1}{C_{ci}} + \frac{1}{C_{\gamma i}}\right)^{-1}$. Within that sum, the contribution from quantum capacitance is $C_{\gamma i} = \gamma_i e^2$ and that from the geometrical capacitance is $C_{ci} = \frac{\epsilon_i}{d}$, with $d = 0.3$ nm as the distance for both cases.[20, 41] The following dielectric constants are used: $\epsilon_{S1813} = 2.54\epsilon_0$, $\epsilon_{PMMA/MMA} = 4\epsilon_0$, $\epsilon_1 = 9.7\epsilon_0$, and $\epsilon_2 = 3\epsilon_0$, where $\epsilon_0$ is the vacuum permittivity.[20, 44-46]

This charge transfer calculation depends on the quantum capacitance parameters $\gamma_1$ and $\gamma_2$, which have been roughly determined for EG devices as $\gamma_1 = 5 \times 10^{12} eV^{-1} cm^{-2}$ and $\gamma_2 = 1.5 \times 10^4 eV^{-1} cm^{-2}$.[20, 41] For the $p$ region, we only need to use a PMMA/MMA spacer layer thickness of 100 nm, yielding a predicted transfer of $\Delta q = 5.9 \times 10^{10}$ cm$^{-2}$ for the corresponding measured hole density of $1.22 \times 10^{11}$ cm$^{-2}$. Subsequently, we repeat the calculation for the $n$ region, bearing in mind that the two polymer spacer layers must be summed in series. The S1813 thickness was 1.2 μm, as per the example AFM images, and the PMMA/MMA thickness slightly increased along the S1813 wall due to the changes in flow during spin coating. The expected charge transfer is $\Delta q = 2.2 \times 10^8$ cm$^{-2}$, a couple of orders of magnitude lower, which makes sense because the observed electron density does not change very much with UV exposure.

Though the capacitance model gave us insight into what charge dynamics to expect in the device, additional information was needed to determine how the charge transfer and total carrier density changed across the junction as a result of the dielectric spacer thickness [see Supplementary Data]. For this, we performed similar coating methods on an example device to gauge whether an identical UV exposure will change the carrier density. For a device having a 300 nm-thick S1813 layer, the electron density remained the same. However, for a device having an approximate S1813 thickness of 41 nm, from reactive ion etching the S1813 layer, the $n$ region changed by nearly one part in $10^{10}$ cm$^{-2}$, but still had a region providing quantized plateaus [see Supplementary Data]. This small shift corresponds to a calculated charge transfer of $1.9 \times 10^9$ cm$^{-2}$, an order of magnitude lower than that of the $p$ region. In other work, a polymethylglutarimide spacer layer of around 200 nm yielded similar results.[32] From these factors, and combining the information obtained from the AFM profile, one can claim that the junction width in these devices have an upper bound of 200 nm, well within the range of junction widths enabling $v = 2$ edge state equilibration.[47] With the knowledge of an upper bound, one can predict the spatial dependence of the carrier density along the *pnJ* by using the charge transfer model. The resulting behavior is shown in Figure 2 (c).

*3.3 Measuring nonconventional fractions of the v = 2 Hall resistance*

With the essential determination that UVP is a viable method for creating large-scale *pnJ*s, we now look to verify expected electrical behavior in a *p-n-p* device.[35] Circuit simulations were also implemented to assist in predicting varied configurations, but first started with the well-known case of injecting a current along the length of the device and measuring various resistances, Hall and otherwise. This verification procedure is presented in Figure 3. The device and its corresponding circuit simulation model are shown in Figure 3 (a). Three voltage measurements are indicated on the device: a single Hall, a *pnJ*, and opposing corners. They are represented by a cyan, green, and blue line, respectively, and an injected current of 1 µA is used. The circuit simulator drawing, on the other hand, elaborates on how the software interprets the problem of current flow. It first assumes that each gray and pink box represents either a counterclockwise (*p* region) or clockwise (*n* region) edge current, strictly for $B > 0$. The simulations can only be performed for a single polarity of magnetic flux density, so to predict the expected resistive behavior for $B < 0$, one may reverse the polarity of the edge current (pink-gray-pink circuit). Equipotential lines are drawn in red, orange, lavender, and blue, to clarify how the potential is expected to behave in the shown orientation.

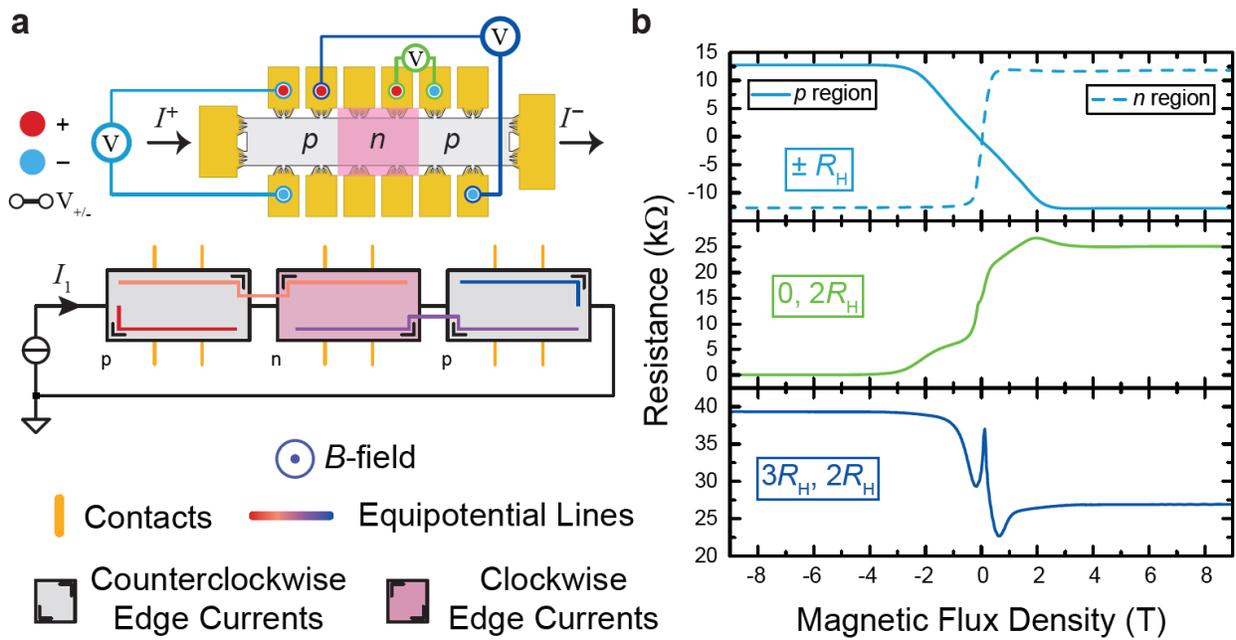

**Figure 3.** (a) Illustrations of the device and the model implemented using LTspice is shown. The voltage measurements are marked with a cyan, green, and blue line connecting the positive (red dot) and negative (cyan dot) terminals, and an injected current of 1 µA is used. Below the device, the current simulator drawing reflects how the software interprets the problem of current flow with the assumptions that each gray and pink box represents either a counterclockwise (*p* region) or clockwise (*n* region) edge current, strictly for $B > 0$. Contacts are shown as gold twigs and equipotential lines are drawn for visual clarity. (b) The resistances of the voltage measurements in **a** are graphed, with the curve colors matching their voltage counterparts in **a**, as a function of magnetic flux density and demonstrate that several multiples of $R_H$ are accessible.

Dividing the current out of the voltage measurements yields our final resistance data shown in Figure 3 (b). According to the model, for positive *B*-field, if one was to measure the resistance from the top to the bottom of the device (cyan voltage measurement), one should expect to see a negative Hall resistance. This result is confirmed in the top panel, with a dotted cyan line showing the experimental result for a Hall measurement in the center of the device, maintaining the positive and negative voltage terminals as the top and bottom, respectively. For the green measurement, one confirms the resistance values of $2R_H$ and 0 for positive and negative *B*-fields, respectively.[35] Lastly, the values of $2R_H$ and $3R_H$ emerge from measuring the opposing corners. These measurements fully verify the functionality of the millimeter-scale *pnJ* device.

The remarkable observations seen in these large *pnJ* devices motivated the exploration of how the overall resistance of the *pnJ* devices can be accurately quantized at other values of $R_H$ under the appropriate conditions. Resistance simulations across points A and B using the circuit simulator (Figure 4 (a)) yielded nonconventional fractions of $R_H$, warranting experimental verification (Figure 4 (b)). Currents were injected at up to three distinct sites on the device, with all currents summing to 1 μA. For general notation, assume that $I_1$, $I_2$, and $I_3$ are positive if they flow from the top to the bottom of the device. If a current is negative, assume that its source has been applied to the bottom of a region, with flow to the top, much like $I_3$ pictured in Figure 4 (a). If one of the three currents is zero, then only two branches are utilized. After the measurement, the overall device resistance is determined from $R_{AB} = qR_H$.

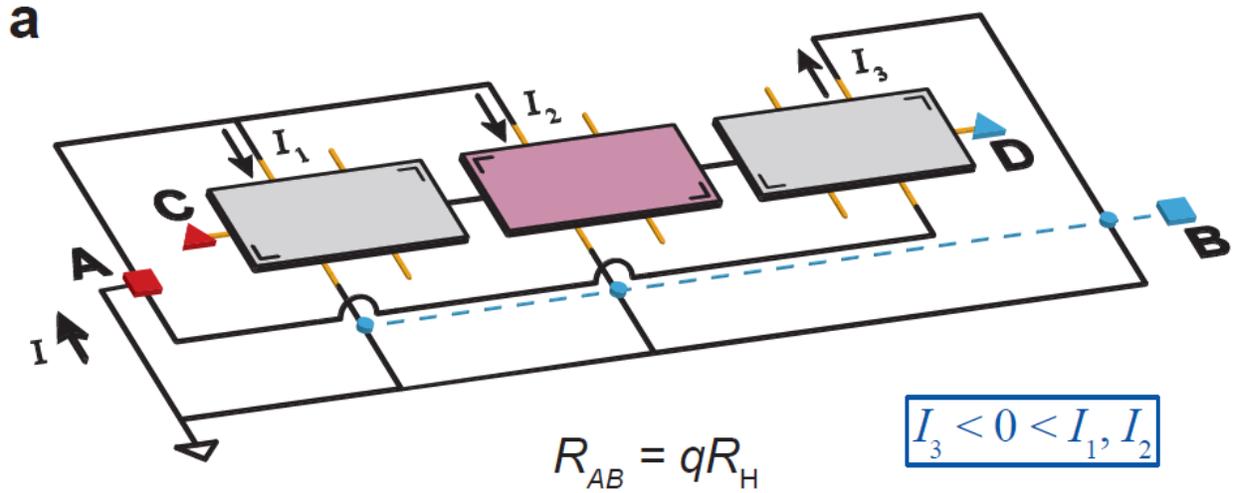

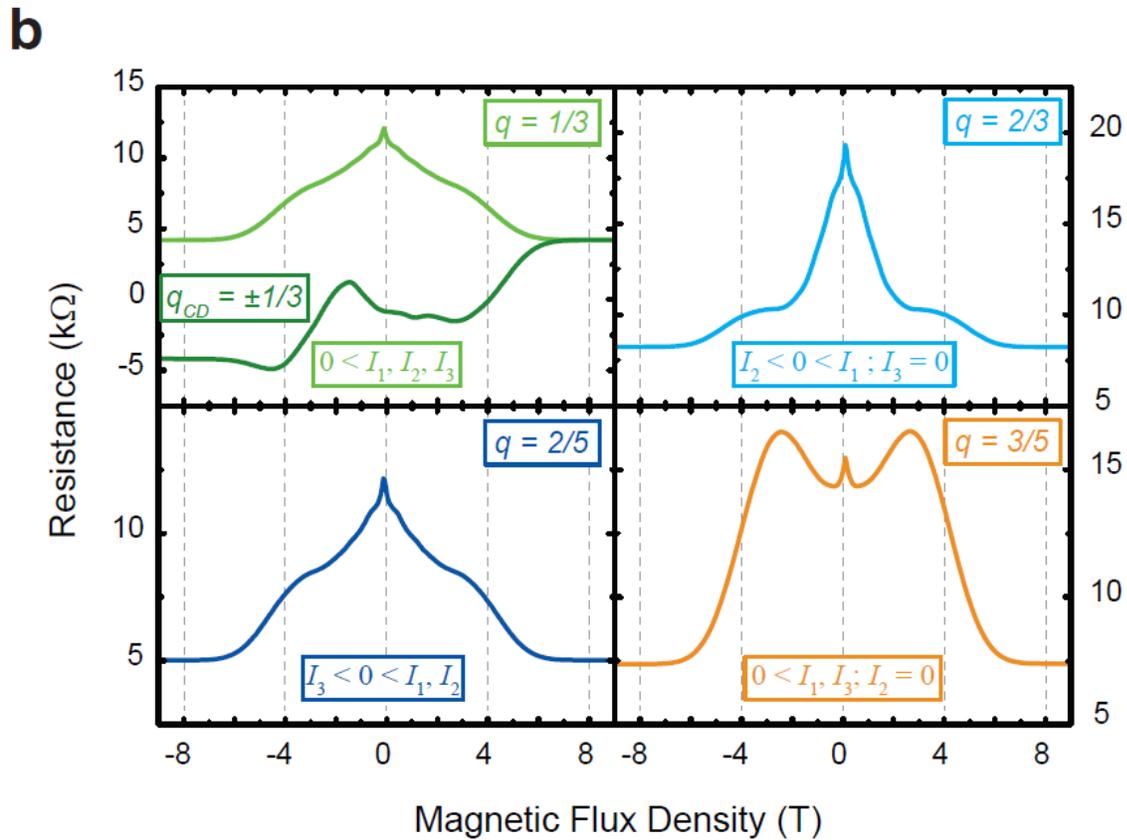

**Figure 4.** (a) An illustration of the circuit model implemented by the spice circuit simulator is shown. A total current ($I$) of 1 μA is used and split among up to three distinct injection points on the device, shown as $I_1, I_2,$ and $I_3$. (b) The measured resistances of several configurations of current injection are plotted as a function of magnetic flux density and demonstrate

nonconventional multiples of $R_H$, including $\frac{1}{3}, \frac{2}{3}, \frac{2}{5}$, and $\frac{3}{5}$. The four measurements are defined by the injected currents, whose positive values correspond to the default flow direction indicated in (a). For negative currents, the connections to the upper side and lower side contacts are switched.

For the configuration $0 < I_1, I_2, I_3$ in Figure 4 (b), $q = \frac{1}{3}$, which may appear to be an intuitive behavior since there are essentially three regions with Hall resistances, albeit with one region having an opposite polarity. Results rapidly become unintuitive when current directions are changed and branches eliminated. The configuration $I_3 < 0 < I_1, I_2$ which is pictured in Figure 4 (a), yields $q = \frac{2}{5}$, revealing that electric equipotentials are substantially redistributed for reversing a single current. Furthermore, when $I_2 < 0 < I_1$ and $I_3 = 0$, a value of $\frac{2}{3}R_H$ is obtained, and for the final displayed configuration of $0 < I_1, I_3$ and $I_2 = 0$ occurs, $R_{AB} = \frac{3}{5}R_H$. The simulations also provide insight on how the effective resistance can switch sign with the polarity of the $B$-field. This antisymmetric behavior can be measured with points C and D, with one example being shown in Figure 4 (b) as a dark green curve in the $0 < I_1, I_2, I_3$ configuration [see Supplementary Data]. In most cases, the accuracy of the deviation of these measurements from their expected values is limited to approximately 1 % due to the elements of the experiment whose uncertainties may also be on that order. Other considerations to make are those of possibly imperfect junctions of insufficient sharpness, whose presence would result in local longitudinal resistances that are unrelated to the propagation of edge states.

When every region of a *pnJ* device displays quantized Hall resistances, but has a different carrier concentration and polarity, the measured resistivity across one or several sets of *pnJ*s depends on Landauer-Büttiker edge state equilibration at the junction.[48-50] In the case of EG, where the Fermi level is typically pinned due to the buffer layer, and where the carrier densities take on values on the order of $10^{11}$ cm$^{-2}$, $\nu = 2$ equilibration becomes most relevant,[6] unlike exfoliated graphene *p-n-p* devices.[51] With this knowledge, one can construct devices with more regions having opposite polarity, like the one shown in Figure 5 (a). Though it can be speculated that these atypical fractions arise from the redistribution of the electric potential throughout the device, this phenomenon is not intuitive with multiple currents in the quantum Hall regime. Because of this difficulty, using a circuit simulator including quantum Hall elements becomes vital for predicting which fractions of $R_H$ are measurable while each region displays the $\nu = 2$ plateau.

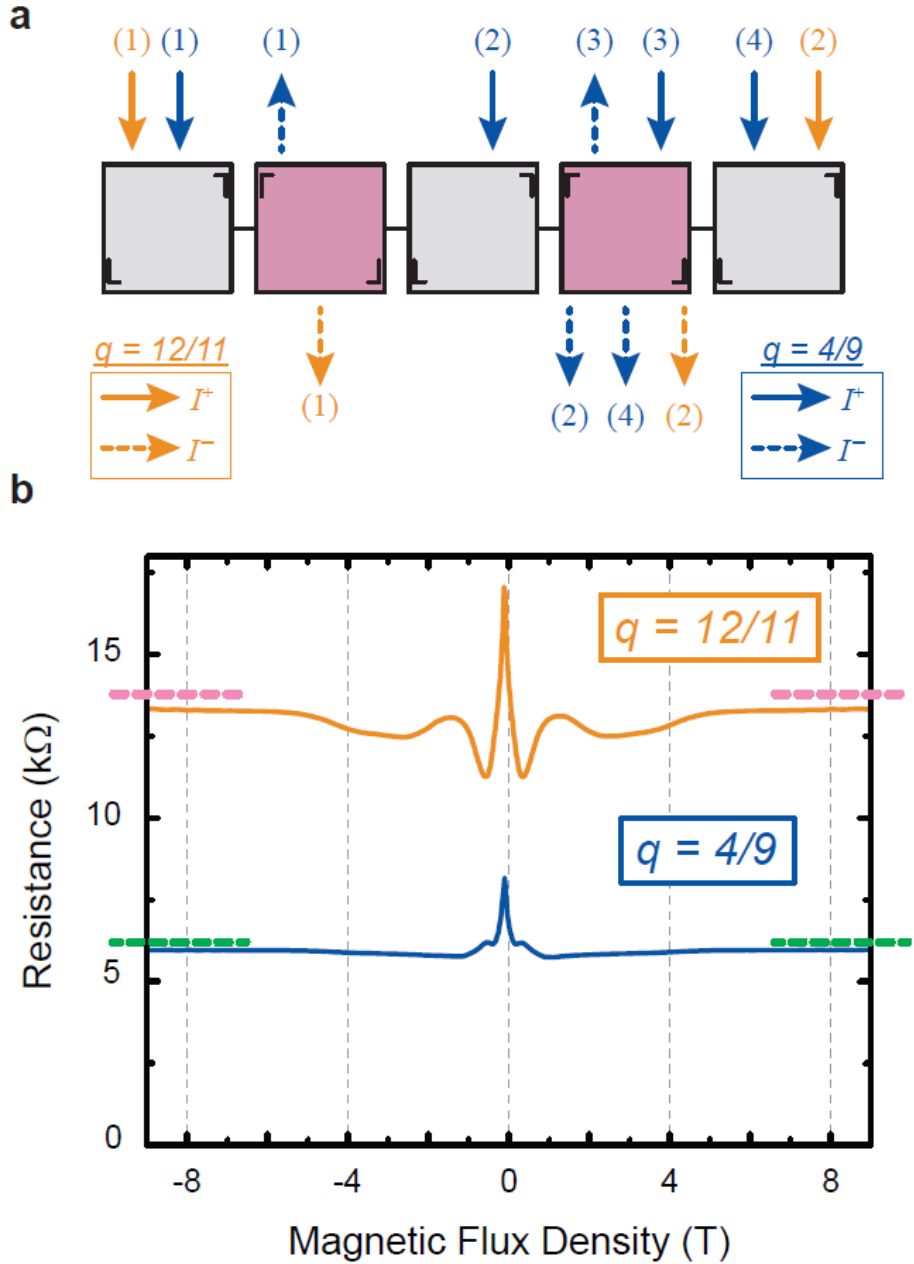

**Figure 5.** (a) An illustration of the circuit model implemented by the spice circuit simulator is shown. A total current ($I$) of 1 μA is used and split among up to four distinct injection points on the device. For the case of obtaining $\frac{12}{11}R_H$, only two currents are injected into the device, shown as solid orange arrows, with corresponding drains as dashed orange arrows. Other fractions, like $\frac{4}{9}R_H$, require four injected currents, indicated by the solid (source) and dashed (drain) blue

arrows. (b) Experimental data are shown supporting the simulated fractions $\frac{12}{11}$ (orange curve) and $\frac{4}{9}$ (blue curve), to within 97 % accuracy. Dotted pink and teal lines are the corresponding simulated values.

In Figure 5 (b), two examples are shown to provide further validation of using the spice circuit simulator for *pnJ* devices. Among the many combinations of input and output currents on the device, the fractional values $\frac{4}{9}R_H$ and $\frac{12}{11}R_H$ are possible results. In the simpler case of $q = \frac{12}{11}$ (orange curve), only two current sources are injected at opposite ends, with the drains exiting from the pair of regions with reverse polarity (see Figure 5 (a)). The data match within 3 %, but more experiments are called for to examine the offset errors, possibly a result of the contact resistances and the lock-in amplifiers used to measure points A and B. Furthermore, an analytical way to determine fractional values for a given configuration is warranted. New *pnJ* devices with many regions in series can now be constructed and explored to find more nonconventional fractions of $R_H$.

## 4. CONCLUSION

To conclude, $\nu = 2$ equilibration was achieved in millimeter-scale *pnJ* devices using only standard ultraviolet photolithography, with junction widths being on the order of 200 nm. Though one group has used a similar process for terahertz applications, they reported neither measured transport properties nor analyzed the reliability of their junction width.[52] The measurements and determinations presented here are crucial for applying them to large scale applications, as well as for other general research that would benefit from having the option of using UVP over the more time-consuming electron beam lithography. This innovative approach of both fabricating *pnJ* devices with UVP and subsequently verifying their functionality begins a new avenue of research in the production of large-scale quantum Hall resistance devices capable of achieving innumerable fractions of $R_H$. Such devices and methods have promising future applicability to both resistance metrology (through scaling the SI ohm) and two-dimensional device fabrication.

## Author Contributions

AFR performed measurements, analyses, and managed the overall project direction. MM performed simulations and measurements. DKP and MK processed samples and performed measurements. AFR, DKP, and MM contributed equally to this manuscript. HMH, HJ, and JH assisted with project concept implementation. ARHW, MO, LC, CTL, and DBN contributed overall project ideas and consulting. All authors have approved the final manuscript.

## Notes

Commercial equipment, instruments, and materials are identified in this paper in order to specify the experimental procedure adequately. Such identification is not intended to imply recommendation or endorsement by the National Institute of Standards and Technology or the United States government, nor is it intended to imply that the materials or equipment identified are necessarily the best available for the purpose. The authors declare no competing interests.

## Funding Sources

All work performed as part of the duties of employees of the United States Government, along with its associated and guest researchers.

ACKNOWLEDGMENT

AFR and HMH would like to thank the National Research Council's Research Associateship Program for the opportunity. The work of DKP at NIST was made possible by arrangement with C-T Liang of National Taiwan University. The work of MM at NIST was made possible by arrangement with M Ortolano of Politecnico di Torino and L Callegaro of Istituto Nazionale di Ricerca Metrologica. The authors thank RE Elmquist and ST Le for fruitful discussions.REFERENCES

[1] A.K. Geim, K.S. Novoselov. The rise of graphene. Nat. Mater., 6 (2007), pp. 183–191.

**2.** A. H. Castro Neto, F. Guinea, N. M. R. Peres, K. S., Novoselov and A. K. Geim. The electronic properties of graphene. Rev. Mod. Phys., 81 (2009), pp. 109–162.

[3] K.S. Novoselov, V.I. Fal'ko, L. Colombo, P.R. Gellert, M.G. Schwab, K. Kim. A roadmap for graphene. Nature, 490 (2012), pp. 192–200.

[4] S. Das Sarma, S. Adam, E. H. Hwang, and E. Rossi. Electronic transport in two-dimensional graphene. Rev. Mod. Phys., 83 (2011), pp. 407–470.

[5] M. Woszczyna, M. Friedemann, T. Dziomba, Th. Weimann, and F. J. Ahlers. Graphene p-n junction arrays as quantum-Hall resistance standards. Appl. Phys. Lett., 99 (2011), pp. 022112-1-022112-3.

[6] T.J.B.M. Janssen, A. Tzalenchuk, R. Yakimova, S. Kubatkin, S. Lara-Avila, S. Kopylov, et al. Anomalously strong pinning of the filling factor nu = 2 in epitaxial graphene. Phys Rev B, 83 (2011), pp. 233402-1-233402-4.22

# Supplementary Data: Nonconventional Quantized Hall Resistances Obtained with ν = 2 Equilibration in Epitaxial Graphene *p-n* Junctions

*Albert F. Rigosi*[†][†], *Dinesh Patel*[†,£], *Martina Marzano*[†,§,⊥], *Mattias Kruskopf*[†,‡], *Heather M. Hill*[†], *Hanbyul Jin*[†,‡], *Jiuning Hu*[†,‡], *Angela R. Hight Walker*[†], *Massimo Ortolano*[§], *Luca Callegaro*[⊥], *Chi-Te Liang*[£], *and David B. Newell*[†]

[†]National Institute of Standards and Technology (NIST), Gaithersburg, MD 20899, USA

[£]Department of Physics, National Taiwan University, Taipei 10617, Taiwan

[§]Department of Electronics and Telecommunications, Politecnico di Torino, Torino 10129, Italy

[⊥]Istituto Nazionale di Ricerca Metrologica, Torino 10135, Italy

[‡]Joint Quantum Institute, University of Maryland, College Park, MD 20742, USA

Contents

1. UV exposure details

2. Modeling and simulation details

3. Spacer layer behavior

[†] Corresponding author: albert.rigosi@nist.gov; 301-975-6572

1. UV exposure details

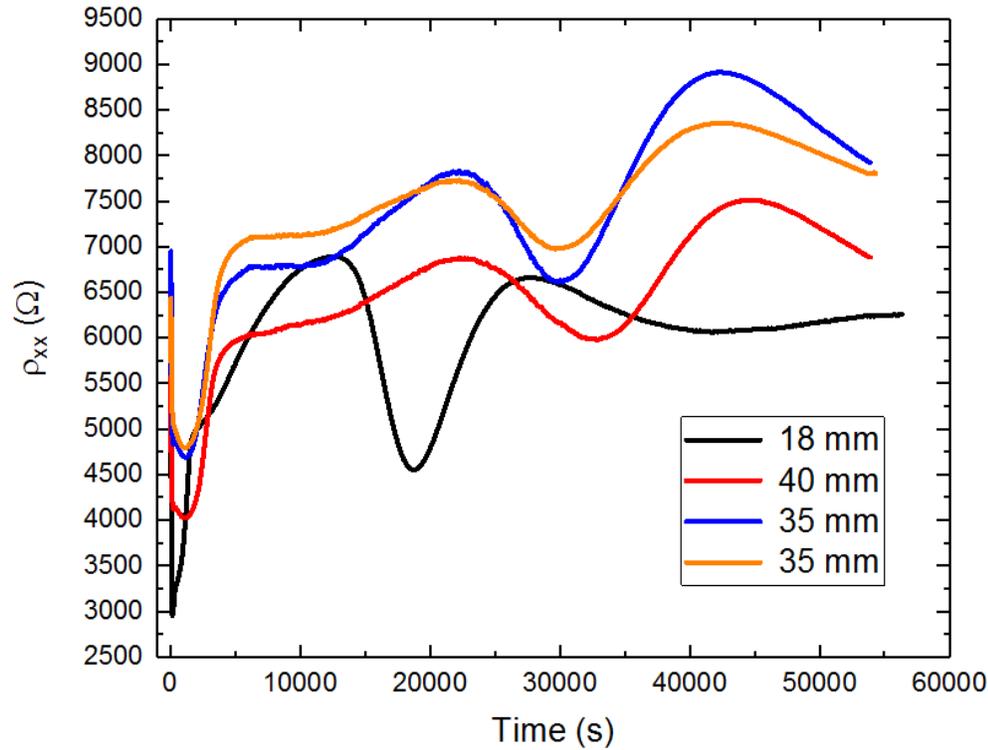

Figure S1. The longitudinal resistivity $\rho_{xx}$ was monitored for several devices being exposed to 254 nm UV light, with some devices being closer or farther from the 17 000 µW/cm² bulb source. The major dip seen at about $1.9 \times 10^4$ s (18 mm), $3 \times 10^4$ s (35 mm for two cases), and $3.3 \times 10^4$ s (40 mm) indicates the start of the transition of an *n*-type region to a *p*-type region, with the Dirac point being successfully crossed after the local maximum approximately $10^4$ seconds after each major dip. Measuring devices with a resistivity at the dip (local minimum) yielded *n*-type behavior in the Hall resistance at 9 T, suggesting that more time was required to convert the *n* regions.

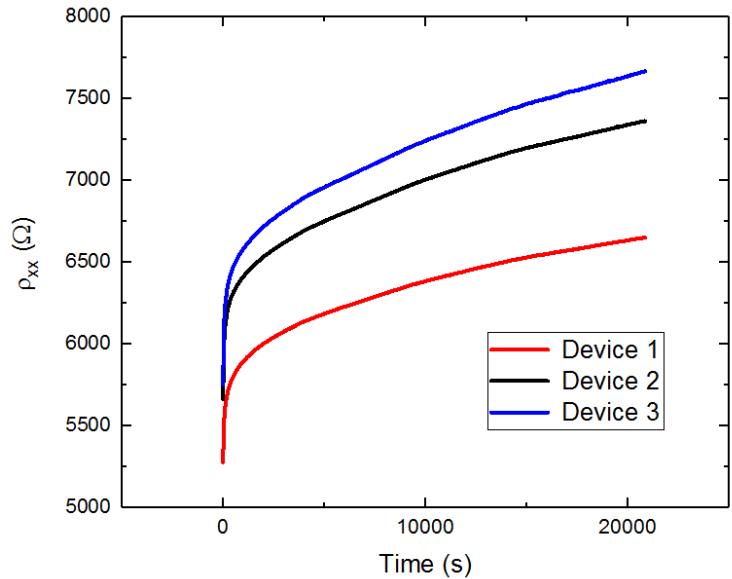

**Figure S2.** The longitudinal resistivity $\rho_{xx}$ was monitored for several devices immediately after exposure to 254 nm UV light. Within $2 \times 10^4$ s, the longitudinal resistivity begins to increase, suggesting the *p*-type region does not permanently maintain its hole density at room temperature and in ambient laboratory conditions.

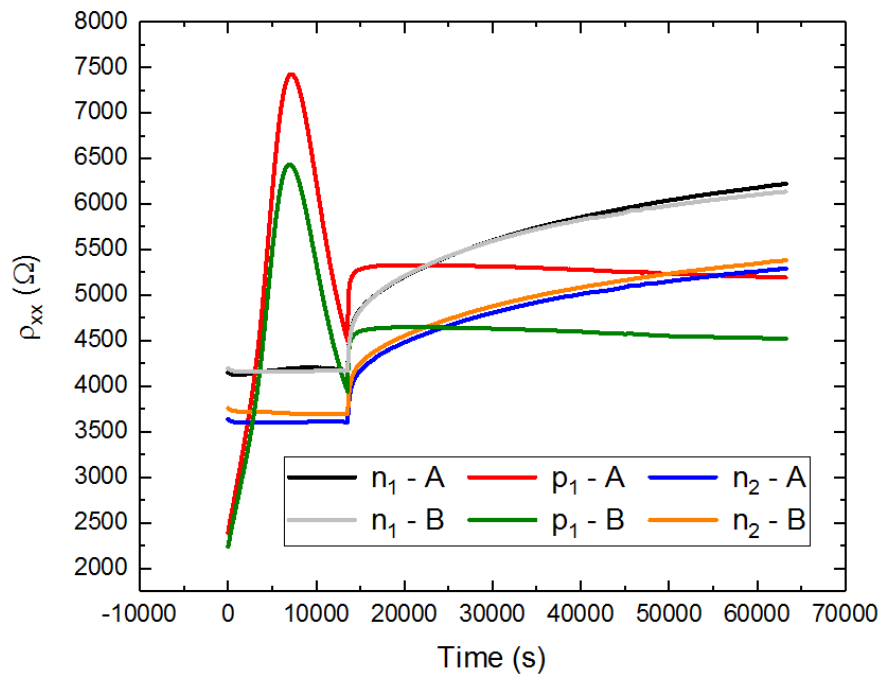

**Figure S3.** The longitudinal resistivity $\rho_{xx}$ at room temperature was monitored as a function of UV exposure time (30 mm away from the source) for an example device which had not been functionalized with $Cr(CO)_3$. Four different regions, intended to remain as *n*-type regions (leading to an *n-p-n* device), are shown as gray and black (two measurements of the first *n* region), and blue and orange (likewise for the last *n* region). The two measurements for the *p* region were in the middle of the device are shown in red and green, and they both show a transient effect within the first $1.5 \times 10^4$ s. At the end of the exposure, all regions maintained an electron density which still provided a quantized Hall resistance at 9 T and 4 K. This demonstrates the necessity of using functionalization to successfully create a *pnJ*.

2. Modeling and simulation details

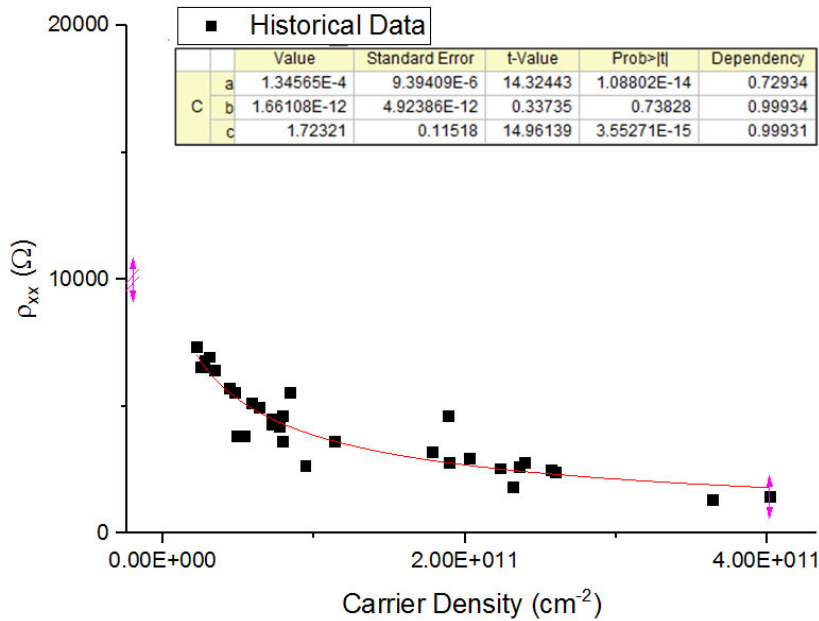

**Figure S4.** To map the relationship between $\rho_{xx}$ to $n_G$, a modified Langmuir fit was used on historical data for the EG devices grown with similar conditions, with the form: $\rho_{xx} = \frac{1}{a+bn^{c-1}}$, where *a, b,* and *c* are fit parameters.

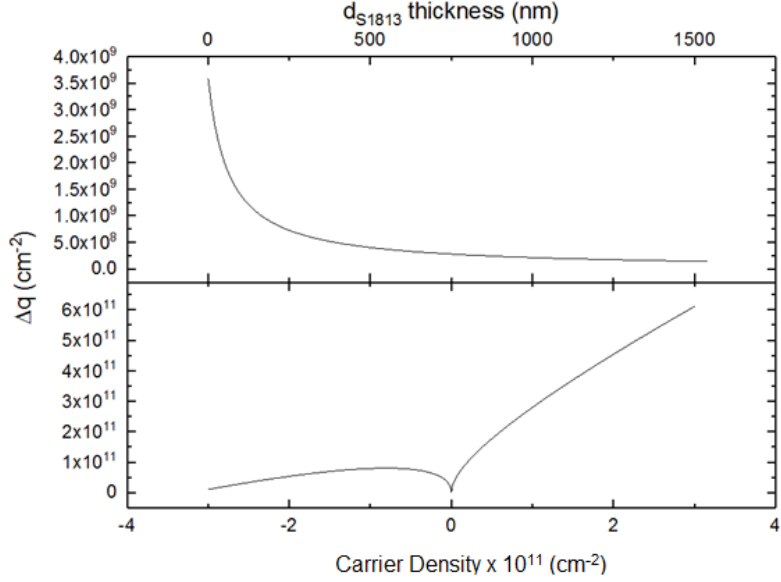

**Figure S5.** The capacitance model in the main text is given by:

$$\frac{C_{\text{poly}}}{e}\left(\frac{e\Delta q}{C_{\text{poly}}} - V_D\right) = \frac{C_{\gamma 2}}{C_{s2}}\left[n_G + (C_{s1} + C_{s2})\frac{E_F}{e^2}\right]$$

Please refer to the main text for all definitions. The charge transferred from EG to ZEP520A is designated as $\Delta q$. The top panels show how $\Delta q$ varies with the thickness of the S1813 layer while keeping the final carrier density in EG constant at $2 \times 10^{10}$ cm$^{-2}$. The bottom panel shows $\Delta q$ as a function of $n_G$ while keeping the S1813 thickness fixed at zero. These trends are shown to elucidate the heavier influence of the final carrier density in EG on $\Delta q$ than the thickness of the S1813 layer.

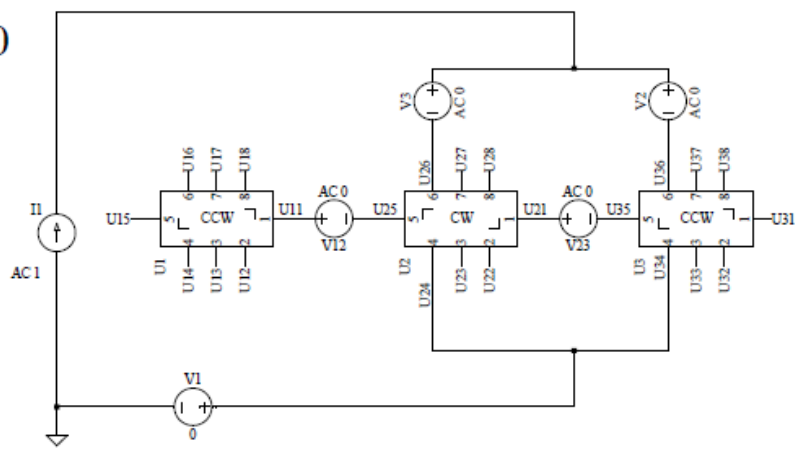

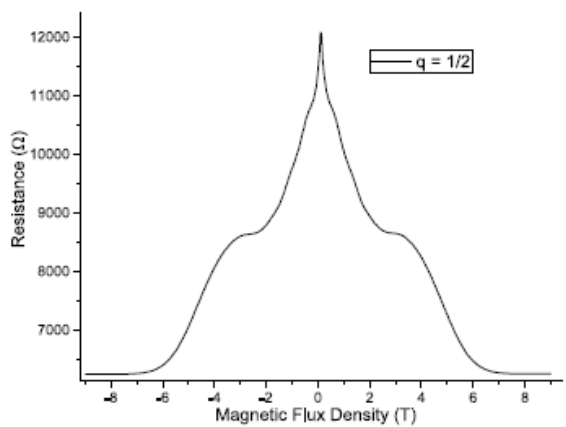 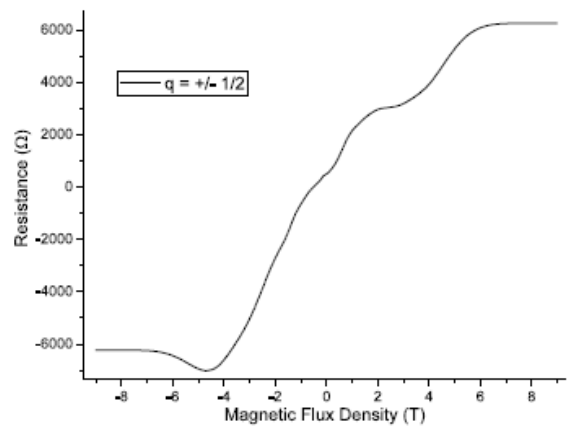

**Figure S6.** To validate the use of LTspice, simulations were performed and subsequently verified by experimental data. This data does not appear in the main text for brevity. The experimental data correspond to the resistance measured across points A and B (left) and C and D (right).

3. Spacer layer behavior

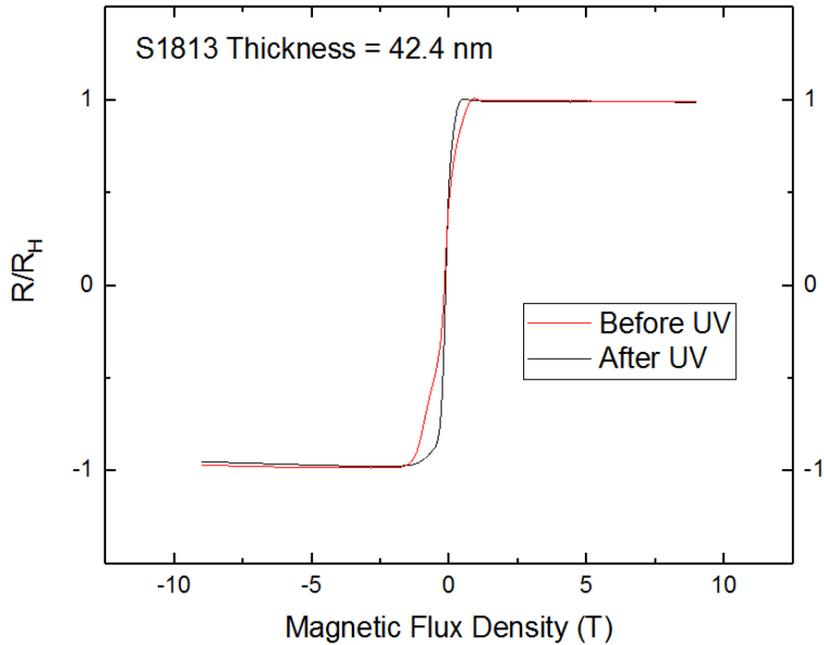

**Figure S7.** The Hall resistance and electron density was checked for three thicknesses of the S1813, two of which are indicated in the main text as sufficient enough a spacer to ensure the stability of the *n* region. To approximate the size of the *pnJ* width, a spacer layer thickness of 42.4 nm was measured with atomic force microscopy, and when that device was exposed to UV light, the regions intended to be maintained as *n*-type experienced a slight change in electron density. Based on the Hall measurements above, the electron density dropped from about $1.7 \times 10^{10}$ cm$^{-2}$ to $0.98 \times 10^{10}$ cm$^{-2}$. The Hall plateau for after UV (in black) was still quantized at 9 T, suggesting that the *pnJ* width had an approximate upper bound of 200 nm (when cross referencing data from AFM profiles in Figure 2 of the main text).

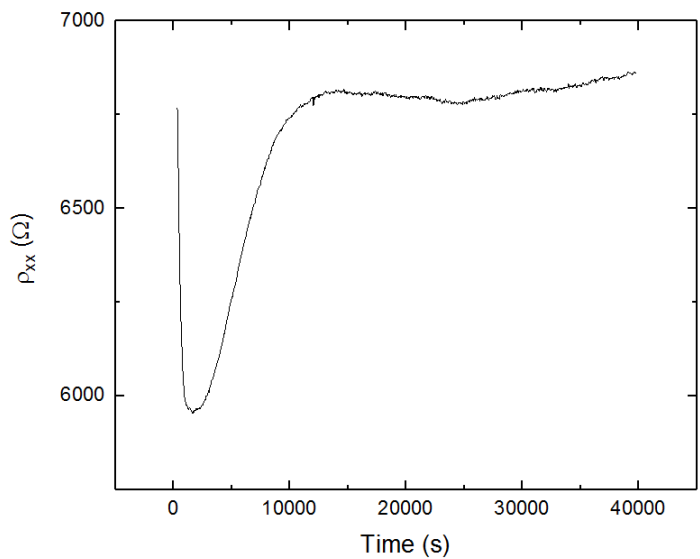

**Figure S8.** The longitudinal resistivity of the device with 42.4 nm thick S1813 was monitored at room temperature between the two Hall resistance measurements in Supplementary Figure 4. The very small change observed in the steady state corresponds well with the small changes observed in the electron density.

NOTES